\documentclass[11pt]{article}

\usepackage{times}
\usepackage{amsmath}
\usepackage{graphicx}
\usepackage{wrapfig,lipsum,booktabs}
\usepackage{subcaption}
\usepackage{amsmath,mathrsfs}
\usepackage{blindtext}
\usepackage{fullpage}
\usepackage{color}

\usepackage{amsmath,mathrsfs}

\begin{document}

\title{Infra-gravity Waves and Cross-shore Transport -\\ A Conceptual Study}

\author{Andreas Bondehagen, Henrik Kalisch\footnote{Henrik.Kalisch@uib.no} \\
\small{Department of Mathematics, University of Bergen, PO Box 7800, 5020 Bergen, Norway} \\
Volker Roeber \\
{\small Universit\'{e} de Pau et des Pays de l'Adour, E2S UPPA, 
chair HPC-Waves, SIAME, Anglet, France}}

\maketitle

\begin{abstract} 
Infra-gravity waves are generally known as small-amplitude waves of periods between $25$ seconds and $5$ minutes. They originate from the presence of wave groups in the open ocean waves and can move freely after being released near the surf zone where they can be further fueled with energy from the spatially varying break point of swell waves (\cite{bertin2018infragravity}). 
As these waves approach the shore, the relative importance of the infra-gravity wave signal increases, and its impact on the shorter waves gets stronger. In addition, infra-gravity waves drive
strong cross-shore currents, which lead to significant back-and-forth motion of the underlying sea water.
This strong cross-shore motion has been made visible by recent field studies.
In \cite{bjornestad2021lagrangian} and \cite{flores2022river},
significant cross-shore movement was detected, and found to be correlated with the infra-gravity wave signal.
In the present work, the connection between infra-gravity waves is explored further using 
linear wave theory and an established numerical nearshore wave model (BOSZ, \cite{roeber2012boussinesq}.
It is shown that in all cases, the presence of infra-gravity waves leads to strong cross-shore motion.
This behavior can be understood by considering the infra-gravity waves as separate free waves,
and then following the fluid particle trajectories excited by these waves. As it is shown,
these trajectories have a very large horizontal extent which
-- if not separated from the main gravity wave field -- appears as a large,
but often not directly visible back-and-forth motion, underlying the more readily observable gravity ocean waves.
\end{abstract}

\section{Introduction}
A sea state can be thought of as a superposition of surface waves of different periods with suitably randomized amplitudes and phase parameters, providing a theoretical description of wave conditions at a particular location in the ocean.
The energy distribution of a sea state is given in terms of a wave spectrum that can often be approximated with simple expressions, which are understood to have fairly broad applicability.
Commonly used expressions include the JONSWAP spectrum \cite{hasselmann1973measurements} for typical North Sea waves, its extension to shallow-water, the TMA spectrum \cite{holthuijsen2010waves}, or the Pierson-Moskowitz \cite{pierson1964proposed} spectrum suitable for the description of fully developed open-ocean swells.
As the individual waves that compose theses spectra propagate through the ocean, fluid particles move in tandem with the waves, 
but at a slower pace, with the particle velocity typically being only a fraction of the wave celerity. 
In the linear approximation, fluid particles trace out nearly circular to elliptic
orbits that do not effectively lead to a mass displacement except for a small forward drift, commonly known as the Stokes drift
\cite{stokes1847theory, lamb1924hydrodynamics, kundu2015fluid}.
While the Stokes drift is the main driving force for wave-induced mass transport in the open ocean \cite{kenyon1969stokes,mcwilliams1999wave},
in the nearshore and in particular in the surf zone, wave breaking is the dominant mechanism for mass transport, affecting processes such as undertow \cite{svendsen1984wave}, surf beat, circulation patterns \cite{davidson2019introduction}, and rip currents \cite{castelle2016rip}.

While ordinary wind-generated gravity waves have periods of $1$ sec to about $30$ sec \cite{munk1951origin,kinsman1984wind},
{\em Infra-Gravity} (IG) waves are ocean waves of usually much larger periods, well above $25$ seconds. Often, waves in the $30$- to $300$-second range are attributed to the IG spectrum; though very long IG waves with periods of up to $15$ min have been reported through observations and numerical modeling efforts (\cite{azouri2016observations}). 
While gravity waves are generated by wind forcing, IG waves appear due to secondary generation mechanisms.
In fact, there are two types of IG waves, bound IG waves, connected to wave groups originating from the deep open ocean, and free IG waves, generated in and around the surf zone.
Regarding bound IG waves, recall that it is well established that ocean waves conventionally appear in groups or sets
(see \cite{longuet1984statistical, thompson1985wave}) 
and that these groups carry long-period oscillations of the mean water level based on the mechanisms outlined in \cite{hasselmann1962non}. These oscillations of the mean water level are known as bound IG waves. Once these bound waves enter shallow water, they are released and propagate freely. This is usually happening close to the break point in shallow water where the waves' group speed depends less on the frequencies of the individual swell waves, but increasingly on the local water depth (\cite{baldock2012dissipation}. In addition, the horizontal movement of the break point location can contribute to the energy in the IG wave band and pointed out by \cite{symonds1982two}. This is essentially based on the fact that individual nearshore waves exhibit varying heights and periods so that wave breaking occurs over a range of different depths and, consequently, at slightly different distances from shore. The concept of the break point mechanism can probably be pictured analogous to a paddle from a wavemaker in a laboratory wave basin. 
Though mostly dominated by the local water depth, free IG waves are free long waves in the sense that they are only governed by the dispersion relation for surface water waves,
which was shown experimentally by Tucker \cite{tucker1950surf}.
IG waves in the surf zone explain the well known surf 
beats observed in \cite{munk1949surf, tucker1950surf}, and 
are connected to various nearshore and coastal processes such as rip currents, run-up/overtopping, 
as well as beach and dune erosion
\cite{russell1993mechanisms,VANTHIELDEVRIES20081028,roeber2015destructive,castelle2016rip,bertin2018infragravity}.

Recent field measurement have revealed the importance of IG waves for several nearshore phenomena
such as river plume oscillations \cite{flores2022river},
net movement of particle tracers \cite{bjornestad2021lagrangian},
and sediment movement \cite{MENDES2020104026}. 
While it has been observed experimentally that cross-shore transport oscillations
correlate strongly with IG wave periods,
there does not seem to exist a firm theoretical explanation of this fundamental process.

The present study aims to provide a theoretical underpinning of how
IG waves drive cross-shore currents.
The basic mechanism can be observed in a simple - yet fundamental - linear analysis of the water-wave problem.
In fact, it is shown that waves of minute amplitude but very long period can cause large horizontal
excursions of the particles in the underlying fluid. This effect can be observed for monochromatic
waves as well as a superposition of wave modes defined by a wave spectrum. It will be revealed
that IG components always dominate the lateral movement of the fluid particles.

For a more realistic approach allowing for nonlinear interactions and wave breaking, the 
nearshore wave model BOSZ, a numerical model of Boussinesq-type, is used \cite{roeber2010shock}.
We set up an idealized beach and apply a wave signal based on an empirical JONSWAP spectrum through boundary forcing.
As in the linear case, it is observed that wave-induced currents in the nearshore are mainly driven by the IG wave signal.
\begin{figure}
    \includegraphics[width=0.98\linewidth]{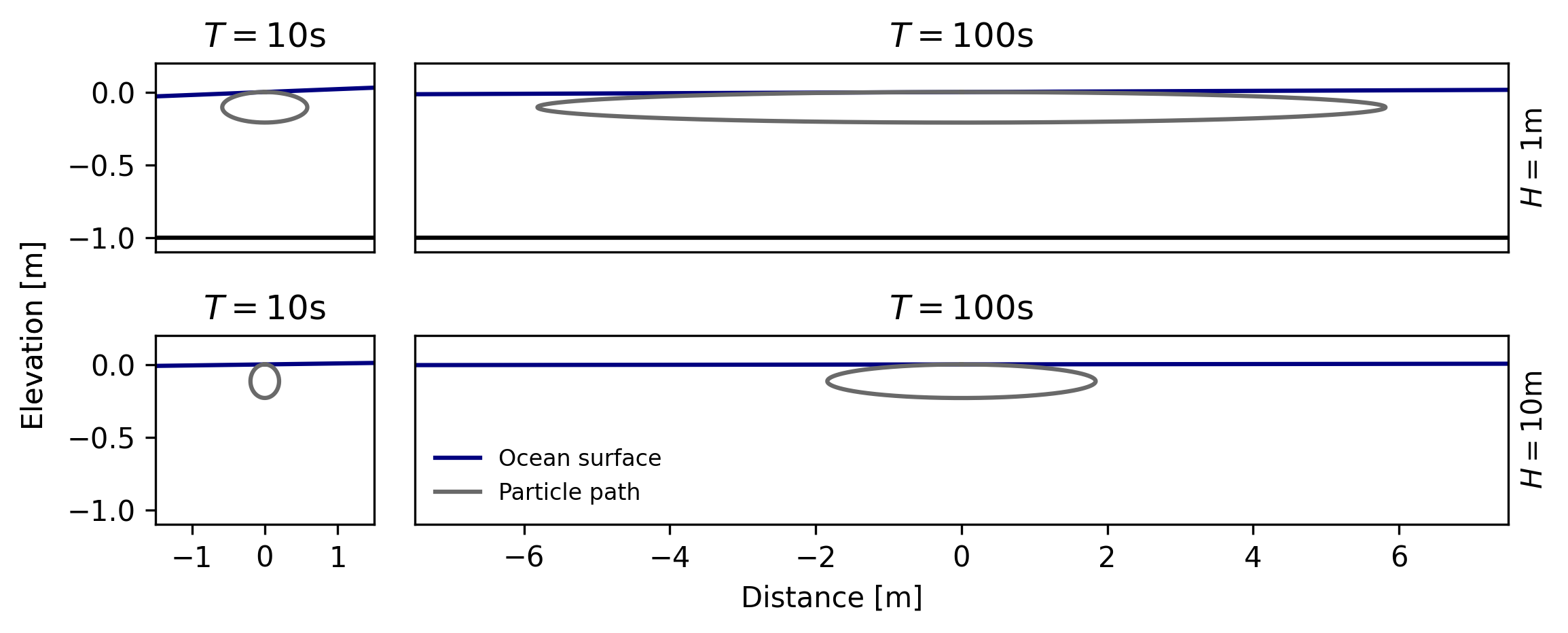}
    \caption{\label{fig:pathLine}
     {\small Pathlines for fluid particles for one period of a monochromatic wave with $0.1$ m amplitude.
     Particle trajectories are found by solving equation (\ref{eq:fluidPart}) with RK4. 
     Left panels: a $10$-second wave with amplitude for depth $H=1$m and $H=10$m. 
     Right: a $100$-second wave with the same amplitude and for the same two depths. 
     It is apparent that the infra-gravity wave induces a much larger extent of horizontal movement.}}
\end{figure}

\section{Linear Theory}
Consider a single wave component in a fluid of depth $H$.
For a monochromatic wave with the free-surface excursion given by $\eta (x,t)=a \cos(kx-\omega t)$, 
the velocity potential is obtained from the linearized free-surface Euler equations as
\begin{equation}
\phi (x,z,t) = \frac{a \omega}{k} \frac{\cosh[k(H+z)]}{\sinh(kH)} \sin(kx-\omega t).
\end{equation}
Here $a$ is the wave amplitude, $k = 2 \pi / \lambda$ is the wave number,
$\lambda$ is the wavelength, $\omega = 2 \pi / T$ is the radial frequency,
and $T$ is the wave period.
The velocity field is obtained by taking spatial derivatives of $\phi$
so that we have
\begin{equation} \label{eq:uxAnalytical}
\phi_x = \frac{a \omega \, \cosh[k(H+z)]}{\sinh(kH)} \cos(kx-\omega t),
\end{equation} 
\begin{equation}
\phi_z = \frac{a \omega \, \sinh[k(H+z)]}{\sinh(kH)} \sin(kx-\omega t),
\end{equation} 
for the horizontal and vertical component, respectively.
Particle paths given parametrically by $(\xi(t), \zeta(t))$
can now be computed by solving the system
\begin{equation}\label{eq:fluidPart}
\begin{aligned}
\frac{d\xi}{dt} = \phi_x(\xi,\zeta,t) = a \omega \frac{\cosh[k (H+\zeta)]}{\sinh(kH)} \cos(k \xi - \omega t) \\
\frac{d\zeta}{dt} = \phi_z(\xi,\zeta,t) = a \omega \frac{ \sinh[k(H+\zeta)]}{\sinh(kH)} \sin(k \xi-\omega t).
\end{aligned}
\end{equation}

Assuming that the particle position stays close to the center $(x_0, z_0)$
allows replacement of the position  $(\xi(t), \zeta(t))$ on the right hand side by the
center position, leading to closed elliptic orbits of the form
\begin{equation} \label{eq:xPathAnalytical}
\begin{aligned}
\xi = x_0 -a \frac{\cosh[k (H+z_0)]}{\sinh(kH)} \sin(k x_0 - \omega t), \\
\zeta = z_0 +a \frac{\sinh[k (H+z_0)]}{\sinh(kH)} \cos(k x_0 - \omega t).
\end{aligned}
\end{equation}
The total extent of horizontal movement of a particle due to a single wave is
then $L(a, k, z) = 2 a \frac{\cosh[k (H+z_0)]}{\sinh(kH)}$.
Note that the second-order approximation of the paths yields a net movement in the direction of the waves, the \textit{Stokes drift} $\bar{u}_L$, already alluded to in the introduction.
The Stokes drift during one wave cycle is given by
\begin{equation} \label{eq:stokesDrift}
x_L = T \bar{u}_L = a^2 \omega k T \frac{\cosh[2k (H+z_0)]}{2\sinh^2(kH)},
\end{equation}
and in shallow water, this can be approximated by the expression
$x_L \sim a^2 \lambda / 2 H^2$. 
It can be seen that in the nearshore zone where the shallow-water approximation
is valid, the ratio between these two quantities is
\begin{equation*}
\frac{x_L}{L(a, k, z)} \sim \frac{\pi}{2} \frac{a}{H}.
\end{equation*}
Since IG wave amplitudes are usually only a small fraction of the water depth
(except during severe sea conditions),
$L(a, k, z)$ dominates substantially over the Stokes drift,
and we neglect the Stokes drift in the present section. However, in a general situation,
a wavefield will feature both IG and gravity wave components,
and the Stokes drift for the gravity-wave components will be considered later.

\begin{figure}[h!]
\includegraphics[width=\linewidth]{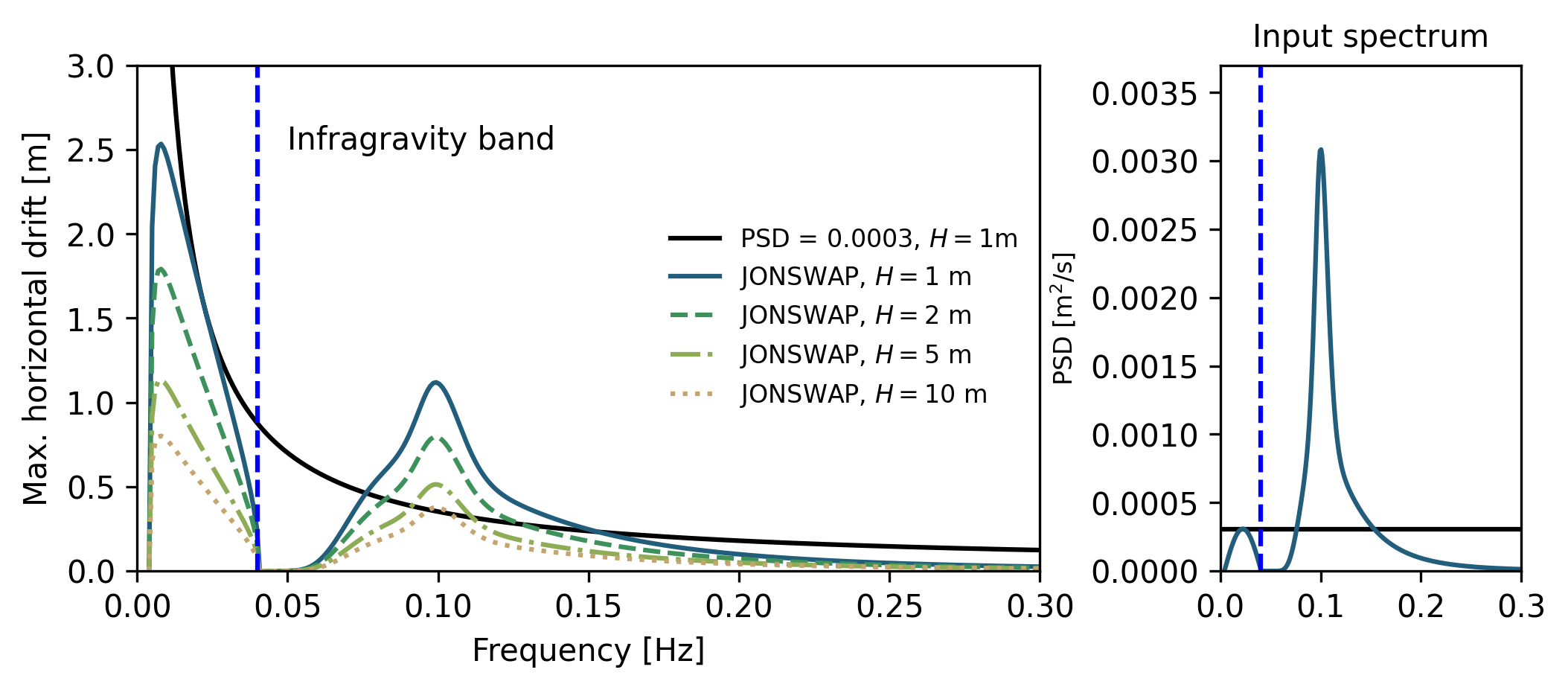}
\caption{\label{fig:analytical} 
         {\small Horizontal extent of particular transport based on analytical solutions of the movement 
         of the particle in a surface wave of a single frequency. The black curve in the left panel shows 
         the horizontal movement associated with different wave components with equal amplitude
         $a=0.025$m. The remaining curves show the maximum horizontal transport due to linear waves with amplitudes chosen from a JONSWAP spectrum with $H_s = 0.05$m and $T_m=10$s, 
         with an added infra-gravity component with $10\%$ of the peak energy, and with varying depths. 
         The right panel shows a representation of the power spectra for the case of a single wave (black line)
         and for the case of a JONSWAP spectrum with additional infra-gravity component (blue curve).}}
\end{figure}

In Figure \ref{fig:pathLine}, particle trajectories associated to surface waves are shown
for four different parameter combinations.
It is apparent that the extent of horizontal movement is much larger for IG waves than for
gravity waves (in the figure, we compare waves of period $10$ seconds and $100$ seconds).
is shown. 
Comparing the upper and lower panels in Figure \ref{fig:pathLine} shows that
the difference in the horizontal extent of the particle movement diminishes with larger depth.
Plotting the total extent of the horizontal movement for waves of in the same depth, with the same amplitude but different periods
yields the black curve in the left panel of Figure \ref{fig:analytical}. 
The plot also shows the movement associated with a superposition of linear waves 
from a JONSWAP spectrum with an added small infra-gravity component. 
The right panel shows the spectrum, and the green and dashed curves in the left panel show
the extent of horizontal movement at difference depths.
For both these examples the infra-gravity waves dominate the movement by far. 

%
%
%


Using linear wave theory allows adding an arbitrary number of wave components in the form
\begin{equation}
\eta_i(x,t) = A_i \cos(k_i x - \omega_i t + \phi_i),
\end{equation}
where the amplitude $A_i$ is Rayleigh distributed and the phase $\phi$ is uniformly distributed.
The free surface is then written as a superposition $\eta(x,t) = \sum_i \eta_i$ and
the fluid velocity at the free surface is given by
\begin{equation}
u(x,z,t) = \sum_i \omega_i A_i \frac{\cosh(k_i (H + z))}{\sinh(k_i H)} \cos(k_i x - \omega_i t + \phi_i)
\end{equation}
and
\begin{equation}
v(x,z,t) = \sum_i \omega_i A_i \frac{\sinh(k_i (H + z))}{\sinh(k_i H)} \sin(k_i x - \omega_i t + \phi_i).
\end{equation}
The movement of a fluid particle can then be described by a coupled system of differential
equations similar to \eqref{eq:fluidPart} as
\begin{equation}\label{eq:fluidPartspectrum}
\begin{aligned}
\frac{d\xi}{dt} = u(\xi,\zeta,t), 
\frac{d\zeta}{dt} = v(\xi,\zeta,t).
\end{aligned}
\end{equation}
Since the expressions for $u$ and $v$ are known explicitly, these equations can be solved
with a standard numerical solver, such as a Runge-Kutta scheme.
Instead of assuming that the velocity of the particle is close to that near the original center, we compute the particle paths directly which is feasible as long as the expressions for $u$ and $v$ are known in closed form (for a similar process applied to cnoidal waves solutions,
see \cite{borluk2012particle}).

\begin{figure}
\includegraphics[width=\linewidth]{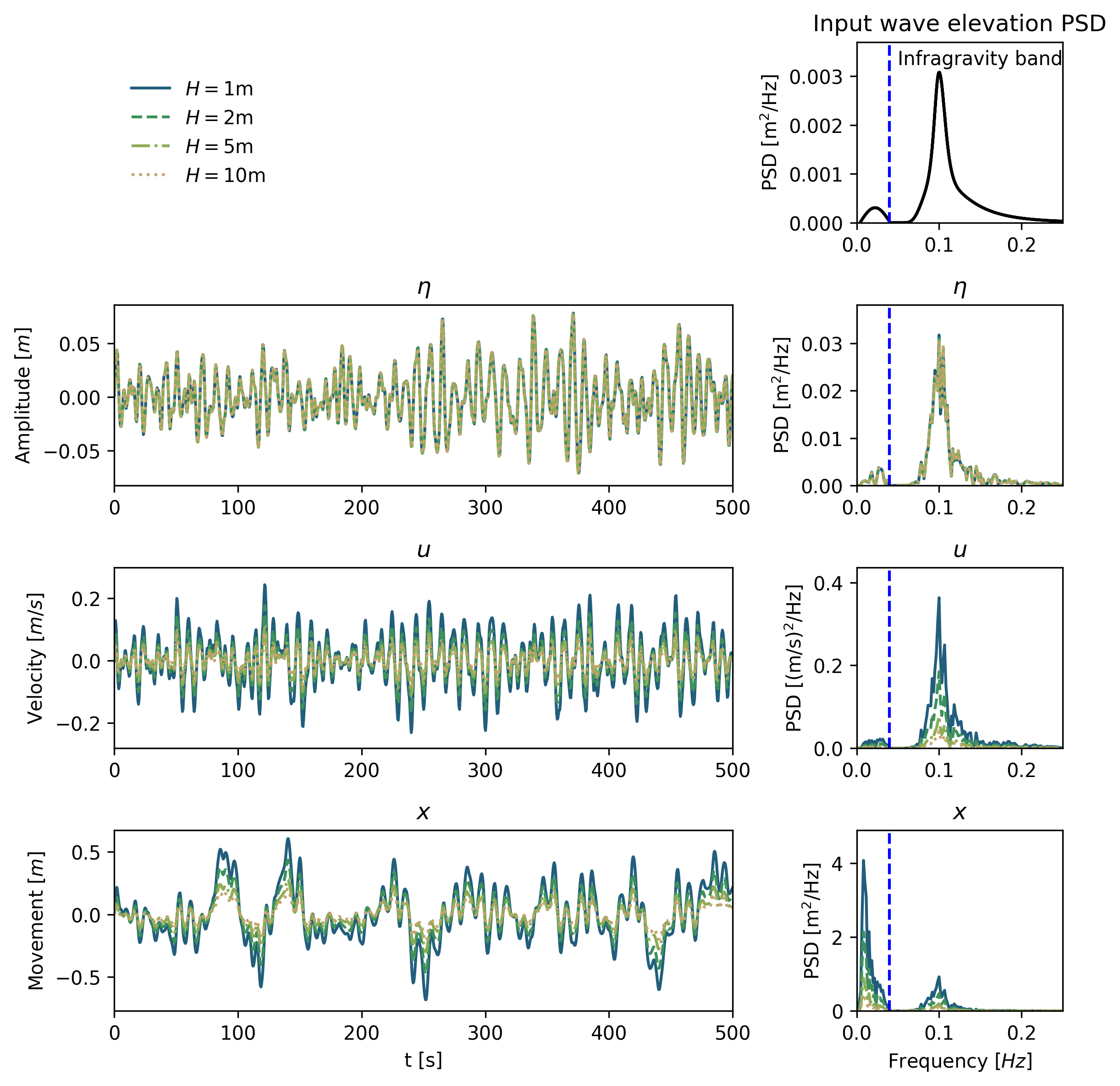}
\caption{\label{fig:experimental}
{\small Experimental results from following a fluid particle moved by a superposition of waves. Top: Amplitudes of waves encountered by the particle. Middle: Velocity experienced by the particle. Bottom: Position of particle over time. Left: Time series of values. Right: Power spectral density of the time series. Colors indicate simulation depth. Blue 1m, orange 2m, green 5m, red 10m. }}
\end{figure}

For the computations shown in Figure \ref{fig:experimental},
we use $500$ wave components with frequencies between $0.004 Hz$ and $0.5 Hz$.

Following the particle for $1800$s with a time discretization of $0.25$s, using the classical four-stage Runge-Kutta scheme to evolve in time, and removing the linear Stokes drift, we get the results from Figure \ref{fig:experimental}. Notice here that the position of the tracer is dominated by the components in the infra-gravity spectrum 
to a larger degree than predicted by the analytical case, 
while the amplitudes and the velocity are reminiscent of the input spectrum. 

Together with the analytical results, this shows that IG waves have a much larger influence on particle transport and dominate that of higher frequency waves, although they carry only a small part of the total energy.

\section{Nonlinear model}
We next pursue to show that the results detailed above also hold for realistic conditions
in the nearshore. 
To this end, we utilize BOSZ, a phase-resolving nearshore wave model based on the Nwogu equations
\cite{roeber2010shock}, which has been shown to yield accurate results in various situations.
In particular, the model has been compared to both laboratory results \cite{wong2019internal},
data from field campaigns \cite{roeber2015destructive,bondehagen2023wave} and with other models
\cite{lynett2017inter}. The model allows evaluation of wave-driven currents and tracking of 
particles.

\begin{figure}[h!]
    \centering
    \includegraphics[width=\linewidth]{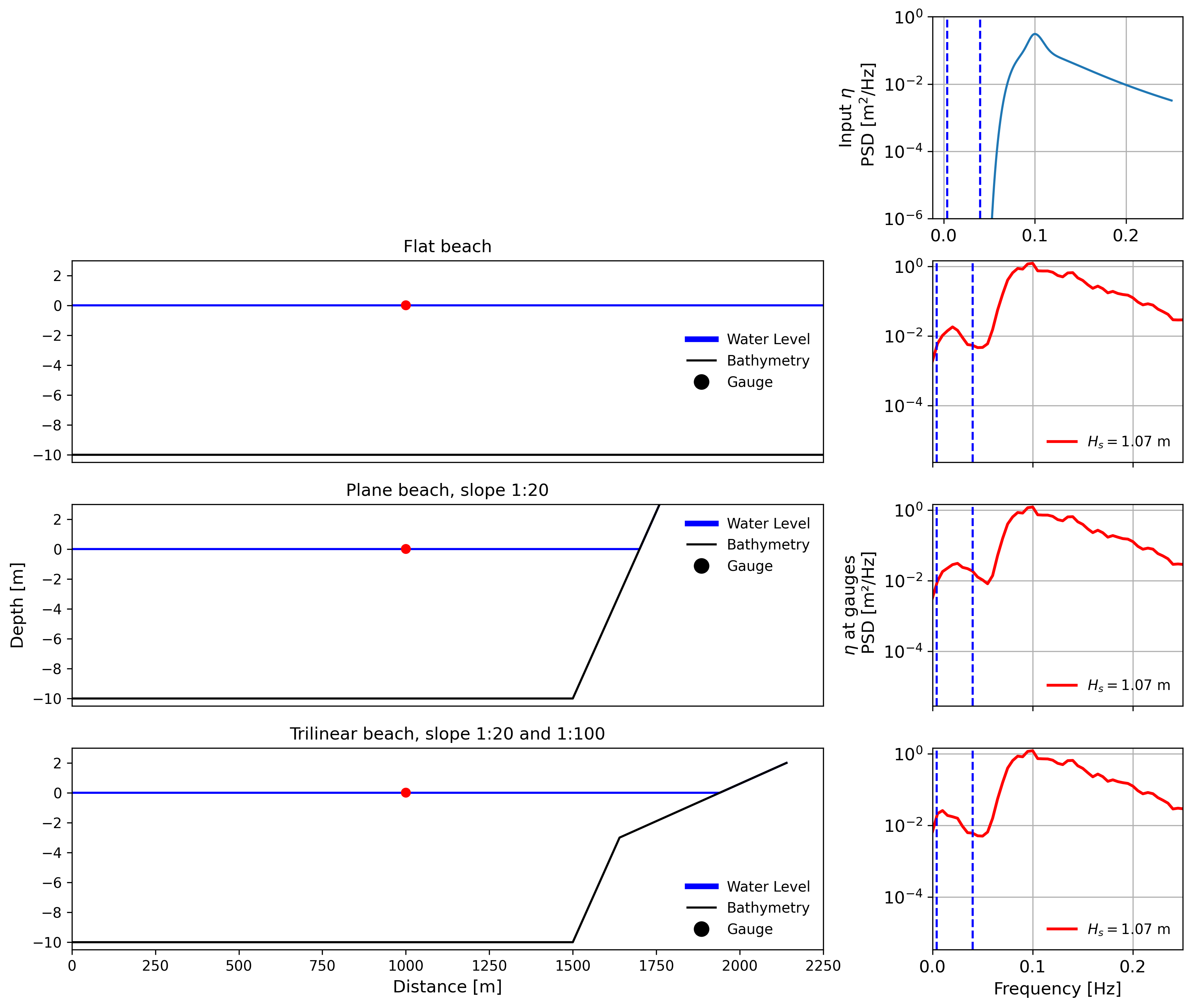}
    \caption{\small 
\label{fig:ig_generation}
Comparison of infra-gravity wave generation from BOSZ. Upper right: the input JONSWAP spectrum to the model with $H_s=1$m and $T_p=10$s. Notice the lack of IG wave components in the input. Second row: A flat beach is simulated, and bound IG waves are generated as anticipated by \cite{longuet1962radiation}. Third and fourth row: with the inclusion of a beach where waves can break the IG band is increasingly significant. This is likely due to free IG waves generated in shallow water around the break point \cite{symonds1982two, baldock2012dissipation}.}
\end{figure}

The model is driven by imposing a sea state from a JONSWAP spectrum near the left boundary.
The sea state is the same as in the earlier tests, but without the infra-gravity component 
(see Figure \ref{fig:ig_generation} upper right panel). In the present case, we first aim to observe IG wave generation due to nonlinear interactions in the governing equations.

To verify that BOSZ can model this IG-generation a computation is run for 3 hours with a 
1 hour ramping time for 3 different bathymetry configurations. The wave's free surface elevation is recorded at $1$ Hz at a virtual gauge in form of a time series, from which the power spectral density is calculated. 
In Figure \ref{fig:ig_generation} one can see the bathymetries and corresponding power spectral densities (PSD). 
In the flat beach case both ends of the domain are padded with a sponge layer so that no reflection takes place. 
Thus IG wave components that are visible in the spectrum must thus be bound IG waves 
as described in \cite{longuet1962radiation, hasselmann1962non}. 
For the other two bathymetries, a plane beach and a trilinear beach, the IG-band is more pronounced by a factor of 2 to 4. 
This is likely due to free IG waves generated around the limit of the surf zone by the break point and shallow water mechanisms outlined in \cite{symonds1982two, baldock2012dissipation}.
The finding that free waves dominate vis-a-vis bound IG waves is also in line with the measurements 
reported on in \cite{smit2018infragravity}.

\begin{figure}[]
\includegraphics[width=0.92\linewidth]{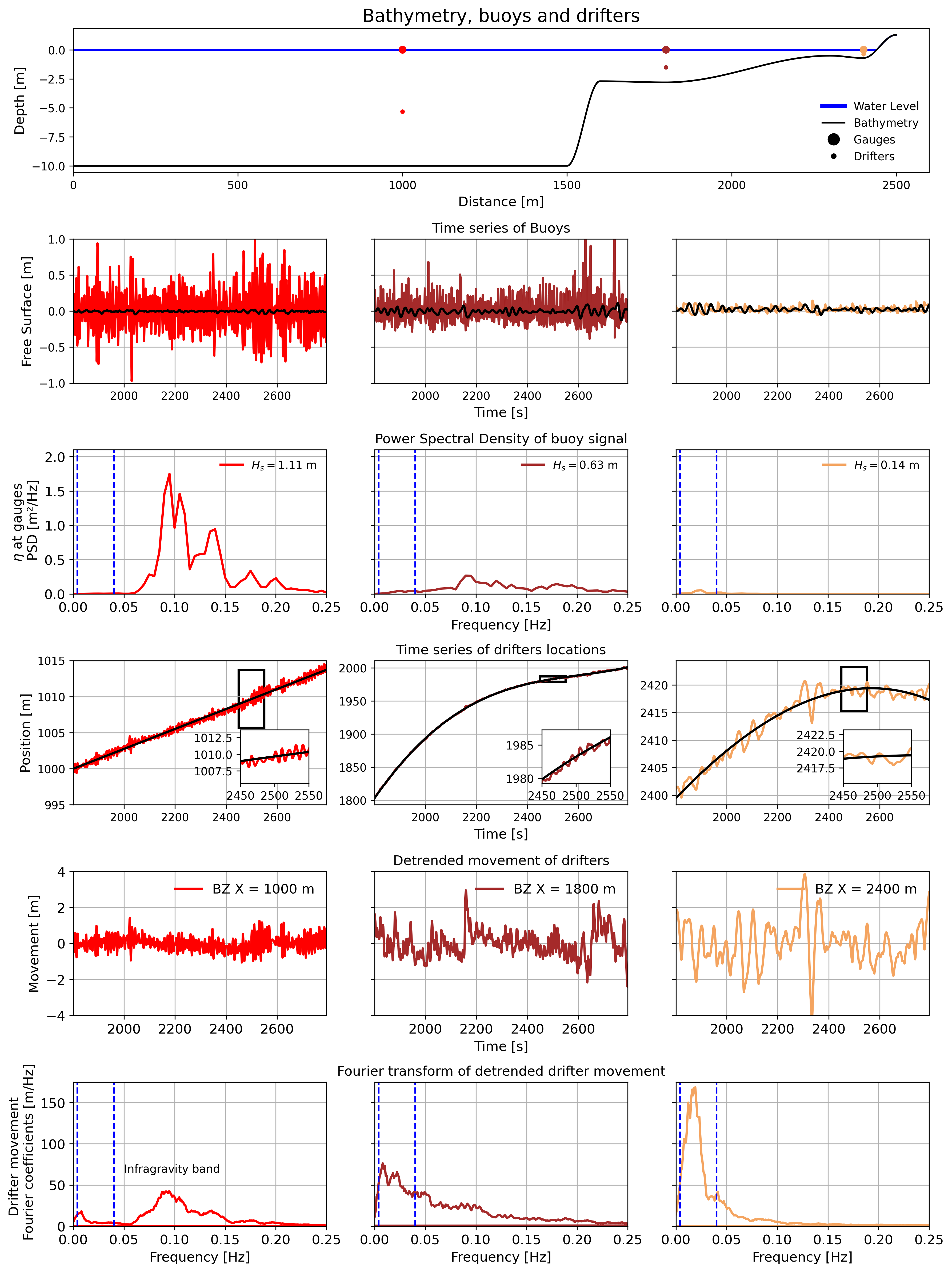}
\caption{\small 
\label{fig:BOSZ_experiment}
Numerical results for particle tracer movements. 
Row 1: Bathymetry and locations.
Row 2: The time series of the wave elevation at the gauges.
The colors and column correspond to the locations in the top plot. 
Row 3: PSD of the wave elevation at the gauges. 
Row 4: Raw time series of the drifters location (color) together with the regression line (black).
Row 5: Time series of the movement about the regression line. 
Row 6: PSD of the movement of the particles after the Stokes drift has been removed.
}
\end{figure}

The bathymetry of the applied test of movement can be seen
in the top panel of Figure \ref{fig:BOSZ_experiment},
with the locations of three measurement points (gauges) and the corresponding tracked drifters marked.
The numerical domain is one-dimensional with a spatial resolution of $DX = 2$m.
This bathymetry features a simple but realistic concept that allows us to
observe the effect of IG waves on wave-induced particle transport. 

The model is run for $3600$ seconds, of which the first $1800$ seconds are dedicated to the full development of the sea state before the fluid particles are introduced. The fundamental frequency detected in a time series analysis at
the wave gauges is $f_0 = \frac{1}{T} \approx 0.001$ Hz, more than enough to reveal the entire IG wave spectrum.
Tracer positions are sampled every $\Delta t = 1$ second, which leads to a Nyquist frequency of $f_{max}=0.5$ Hz, sufficient to track nonlinear wave-wave interactions and observe super-harmonics during shoaling.

Results of an in-depth time series analysis are shown in Figure \ref{fig:BOSZ_experiment}.
In the panels in row 2, the total signal as well as the low-pass filtered data recorded at
the gauges is plotted. Combining this with the PSD as seen in row 3, one can see in the left
column that the JONSWAP spectrum has been correctly generated by the wavemaker,
but with some extra energy in the frequency band at $0.2$Hz. 
In the middle column the waves have travelled over the slope, so that the water depth is now  $2.5$ meters, 
as opposed to $10$ meters in the left column.
It can be seen in the middle plot of row 3 that energy from the ordinary gravity wave spectrum has almost vanished due to bathymetry-driven wave breaking while energy starts to appear
in the IG-band. Continuing on to the rightmost gauge in $1$ meter depth,
both dissipation due to wave breaking, and energy transfer to the 
IG wave continues, and the free surface is now dominated by waves in the IG band. 

In row 4 the position of the three numerical drifters over time is shown. Since the numerical drifters move in the direction of the waves due to Stokes drift, it
is necessary to detrend their time series signal around a moving mean position as shown in row 4 of Figure \ref{fig:BOSZ_experiment}.
For the drifter in $10$ meters depth, the Stokes drift as defined in Eq. \ref{eq:stokesDrift}
is constant due to the flat bathymetry, so that a first-order linear regression can be used to
remove the Stokes drift from the observed drifter path.
For the drifters in the shallower locations, subtracting the Stokes drift is not as straightforward.
As the numerical drifters move to shallower water, the wave field and hence Stokes drift change
yielding a non-constant drift, which makes a linear curve fit problematic.
Nevertheless, a 3rd and 4th-order polynomial provided a much better fit as indicated by the black line.
The movement about the changing mean position for the three drifters can be seen
in row 5 in Figure \ref{fig:BOSZ_experiment}.
Lastly, the Fourier transformation of this signal was taken resulting in the plots of row 6.
Comparing row 3 and 6 shows the importance of the IG band on the particle movement.
In the left column it is impossible, on a linear scale, to see the PSD in the IG band
while it still has a significant influence on the movement of the drifter.
For the middle row it is possible to notice the PSD component in the IG band,
but it is still small compared to the one from the ordinary gravity waves.
Nevertheless, the particle movement is dominated by the IG wave component.
Lastly, in the right plot, next to shore the movement is almost entirely controlled by the IG band.
Also looking at the other particle paths, the increased energy associated with IG waves now yields a much larger movement than earlier even though 
the wave field is much smaller.
\section{Discussion}

In the present analysis, the dynamics of IG waves in the nearshore have been considered in a two-dimensional setting, i.e. along a shore-normal transect. In this simplified situation, a direct link between the IG signal and cross-shore fluid particle movement has been identified.
In practice, the properties of IG waves depend to various levels on other factors such as the direction of the incoming wave field \cite{herbers1995generation}, local beach morphology \cite{bryan1998field}, or the tide stage \cite{melito2022wave}, to name only a few.
Obviously, three-dimensional dynamics will generally affect the IG wave signal, in particular in connection with the
appearance of edge waves \cite{herbers1995infragravity}, IG wave reflection \cite{sheremet2002observations}, as well as eventual propagation off-shore and refraction at the shelf break \cite{smit2018infragravity}.
As already mentioned, IG waves are linked to a wide range of nearshore processes, such as short-wave modulation, 
surf zone circulation, and interactions with the sediment  \cite{bertin2018infragravity,de2016cross,melito2020hydrodynamics}. The results in this paper are in line with recent observations of horizontal movement of freshwater plumes \cite{flores2022river} and also with wave-by-wave motions of tracer particles in the nearshore \cite{bjornestad2021lagrangian}. Both have been reported to correlate with measured IG waves.

\section{Conclusion}

It has been brought to light that infra-gravity waves have a large impact on the horizontal cross-shore transport in fluids. This transport is essentially oscillatory in nature, i.e. it describes a back-and-forth motion of fluid particles. This is in contrast to the Stokes drift of gravity waves that is always acting in the direction of wave propagation. The IG wave-induced motion is large in proportion to the IG wave amplitude. In fact, a small IG wave amplitude can lead to very large back-and-forth motion in the fluid. The phenomenon essentially applies in the same way to a single wave component or a spectrum, and it also occurs in a similar fashion in
nonlinear waves (i.e. as shown by a Boussinesq-type model). The principles also hold consistently for non-uniform sea floors. 
While the mechanism described in this letter
may not be easily detectable in a three-dimensional setting - as it will rarely be the only controlling process, the basic principles provided here suggest that the horizontal motion induced by the IG waves always play an important role. \\\\

{\bf Acknowledgments}
The authors would like to thank Alexander Horner-Devine for providing some inspiration
for undertaking the present study.
Henrik Kalisch acknowledges support from {\em Bergen Universitetsfond}.
Volker Roeber acknowledges financial support from the I-SITE program
{\em Energy \& Environment Solutions} (E2S), the Communaut\'{e} d'Agglom\'{e}ration Pays Basque (CAPB),
and the Communaut\'{e} R\'{e}gion Nouvelle Aquitaine (CRNA)
for the E2S chair position HPC-Waves.


\end{document}